\documentclass[aps,twocolumn,floatfix]{revtex4}%
\usepackage{amsfonts}
\usepackage{amsmath}
\usepackage{amssymb}
\usepackage{graphicx}
\usepackage{graphicx}%
\begin{document}
\title{Transformation of a three-tone phase modulated CW laser field into sequence of
short pulses}
\author{Rustem Shakhmuratov}
\affiliation{Zavoisky Physicsl-Technical Institute, FRC Kazan Scientific Center of RAS,
Kazan 420029, Russia}

\begin{abstract}
This paper presents generation of Fourier-limited pulses from a periodically
phase-modulated CW laser field by phase manipulation of its spectral
components. Phase modulation in the form of modulus function of sine (MFS) is
considered. It allows to produce pedestal-free pulses with controllable
repetition rate and duty cycle. Sidelobes of these pulses can be excluded by
choosing a proper modulation index. The proposed method allows to produce
pulses with 50\% duty ratio. It is shown that three-tone modulation of phase
allows to simulate MFS phase modulation.

\end{abstract}
\maketitle

\section{Introduction}

High-repetition-rate optical pulses are of interest in many applications
related to optical communications, metrology, and signal processing
\cite{Quinlan2009}. Pulse generation by mode-locked lasers is one of the
widespread technics. However, the repetition rate of these lasers is limited.
New microresonators offer sources of short pulses with much high repetition
rate, which varies from 10 GHz to 1 THz depending on the resonator size
\cite{Kippenberg2011,Weiner2017,Kippenberg2018}. However, their repetition
rate is not tunable, as it depends on the mode spacing for a given device. In
addition, such sources require thermal stabilization, and their parameters
critically depend on the pump power, which requires special control.

Meanwhile, tunable generation of high-repetition-rate optical pulses with
arbitrary shape can be synthesized by phase-intensity spectral shaping of
frequency combs \cite{Cundiff2010}. Moreover, such combs with narrow and
stable components can be implemented by a periodic phase modulation of a CW
laser. Examples of phase-intensity spectral shaping of these frequency combs
are discussed in \cite{Mamyshev1994,Weiner2006,Finot2019,Sheveleva2020,
Sheveleva2021}.

Frequency combs of this type have a long history. They were first created by
phase modulating a CW field transmitted through an electro-optical modulator
fed by a sinusoidal voltage. Filtering of this field through a
frequency-dispersive linear delay line results in the formation of optical
pulses with a repetition rate equal to the modulation frequency
\cite{Pearson1975,Grischkowsky1978}. This idea is based on the similarity with
a chirp radar, where a linear frequency chirp of the field in time gives
different delays of its spectral components in a frequency-dispersive delay
line, so that the spectral components generated at the end of the phase change
catch up with the components generated at the beginning of the phase change
\cite{Klauder1960}. Bunching of the field components creates a pulse.

The chirp radar concept is applicable to a linear-in-time frequency chirp or a
quadratic-time dependence of the field phase. A group-delay-dispersion (GDD)
circuit is just appropriate for shaping a pulse from this field, since the GDD
circuit gives a phase change of each spectral component of the comb
proportional to $n^{2}$, where $n$ is the number of spectral component,
counted from the central frequency of the field. This approach was applied in
\cite{Kobayashi1988,Kim1996,Otsuji1996,Murata2000,Komukai2005,Torres2006,Torres2011}
to sinusoidal phase modulation, which converts a narrow single-line CW light
into a frequency comb with narrow spectral components. For example, specially
designed electrooptical modulator is capable to create sideband spectrum of 2
THZ full-width half-maximum \cite{Kim1996}. However, in half-cycle of the sine
function, the field frequency increases with time (up chirp) and, in the next
half-cycle, this frequency decreases (down chirp). Therefore, only in one of
the two half cycles, with a suitable chirp, the GDD circuit generates a pulse,
and in the next half-cycle, with an inappropriate chirp, pulse compression
does not occur and we see only a pedestal with a non-zero field intensity. To
eliminate these pedestals between pulses, it was proposed to filter out
unsuitable half-cycles by amplitude modulating the CW field before phase
modulating it with the same frequency \cite{Otsuji1996}. On the other hand, a
time lens, realized by periodically repeated in time quadratic elements of the
field-phase change, creates pedestal-free pulses after the field propagation
through the GDD circuit without any additional treatment
\cite{Azana2017,Shakhmuratov24}. This is due to the correct linear frequency
chirp caused by the parabolic phase dependence over time in each modulation period.

In this paper we consider another type of phase modulation, which is described
by the modulus function of sine (MFS). The frequency chirp caused by this
phase modulation is only ascending or descending depending on the sign of
modulation index, with sharp breaks between periodically repeated elements.
The GDD circuit is also capable to create pedestal-free pulses from this phase
modulated field, however with small ripples between them. This is due to the
nonlinearity of the frequency chirp in time. We found that the phases of the
spectral components of the frequency comb created by the MFS phase modulation
do not follow a quadratic dependence on the number of spectral component $n$.
They have a bell-shaped dependence on $|n|$, which can be approximated by
Gaussian or Lorentzian functions for moderate values of the modulation index.
Compensation of these phases can be accomplished by processing a line-by-line
spectral phase profile using liquid-crystal modulators
\cite{Weiner2006,Finot2019,Sheveleva2020,Sheveleva2021} or fiber Bragg
gratings \cite{Berger2006}. It is shown that the approximation of the MFS by
three harmonics (three-tone phase modulation) allows one to implement such a
specific phase modulation quite simply. A comprehensive analysis of the
properties of the resulting pulses is given and multiplication of their
repetition rate is proposed.

\section{MFS modulation of phase of the CW Field}

Let us consider a field $E_{M}(t)=E(t)e^{-i\varphi_{M}(t)}$ with a
periodically modulated phase, which is described by the function
\begin{equation}
\varphi_{M} (t)=\Delta|\sin(\pi t/T)|,\label{Eq1}%
\end{equation}
where $\Delta$ is the maximum value of the phase shift (modulation index),
$E(t)=E_{0}\exp(-i\omega_{c}t+ikz)$ is the CW field, $E_{0}$ is its amplitude,
$\omega_{c}$ and $k$ are the carrier frequency and the wave number, $t$ and
$z$ are time and spatial coordinates. The evolution of this phase and its time
derivative, which is the chirped frequency, are shown in Fig. 1(a). The
chirped frequency decreases with time each period and abruptly rises to its
maximum value at the end of the period. As can be clearly seen, this chirp is
nonlinear. \begin{figure}[ptb]
\resizebox{0.4\textwidth}{!}{\includegraphics{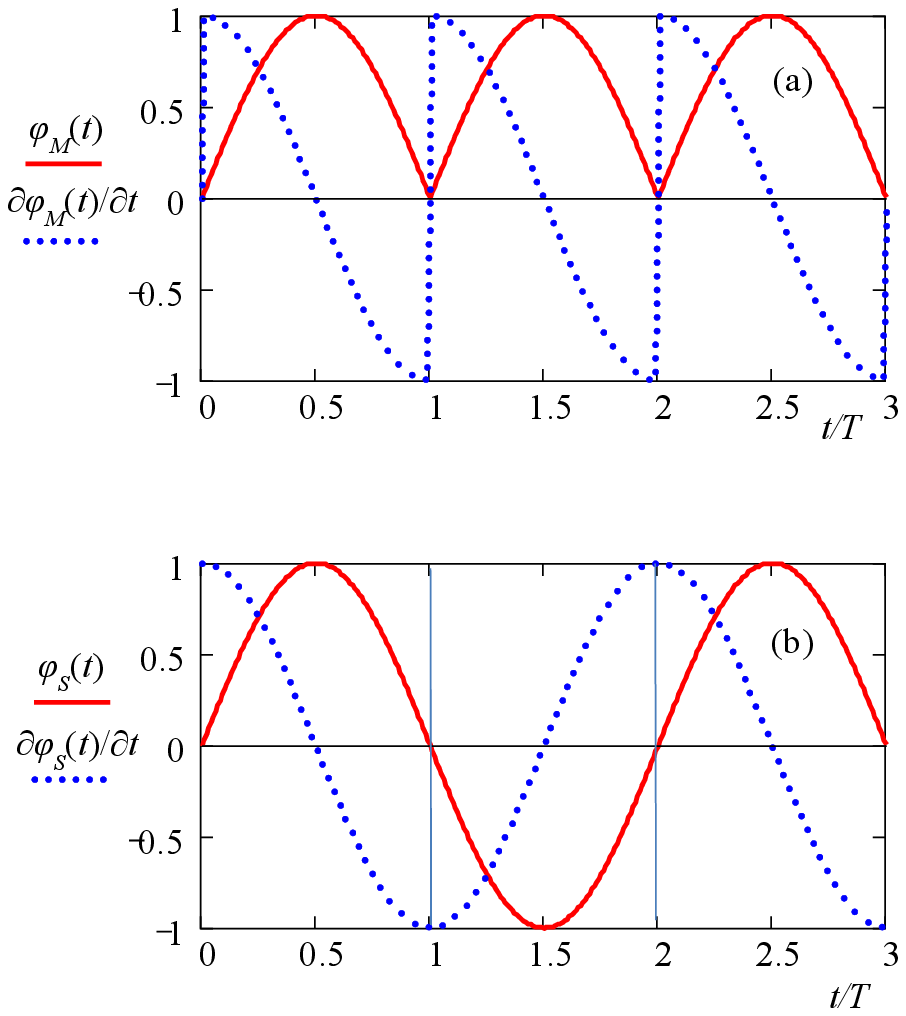}}\caption{(a) MFS
phase modulation $\varphi_{M} (t)$ normalized to $\Delta$ (red line). Dotted
blue line shows the frequency chirp $\partial\varphi_{M}(t)/\partial t$
normalized to $\pi\Delta/T$. (b) Sine phase modulation $\varphi_{S} (t)$
normalized to $\Delta$ (red line). Dotted blue line shows the frequency chirp
$\partial\varphi_{S}(t)/\partial t$ normalized to $\pi\Delta/T$. Vertical blue
solid lines separate domains with opposite chirps.}%
\label{fig:1}%
\end{figure}The sinusoidal modulation of the field phase, $E_{S}%
(t)=E(t)e^{-i\varphi_{S}(t)}$, where
\begin{equation}
\varphi_{S} (t)=\Delta\sin(\pi t/T),\label{Eq2}%
\end{equation}
is shown in Fig. 1(b) by red solid line, and its time derivative is shown by
dotted blue line for comparison. The sinusoidal phase modulation has
down-chirp and up-chirp in each period, which is taken equal to $2T$ for
visual comparison with the MFS.

A Fourier series expansion of the fields $E_{M}(t)$ and $E_{S}(t)$,
\begin{equation}
E_{M,S}(t)=E(t)\sum_{n=-\infty}^{\infty}A^{M,S}_{n}(\Delta)e^{-i2\pi
nt/T_{M,S}},\label{Eq3}%
\end{equation}
gives the amplitudes of their spectral components
\begin{equation}
A^{M,S}_{n}(\Delta)=\frac{1}{T_{M,S}}\int_{0}^{T_{M,S}}e^{-i\varphi_{M,S}
(t)+i2\pi nt/T_{M,S}}dt,\label{Eq4}%
\end{equation}
where $n$ is an integer, $T_{M}=T$ and $T_{S}=2T$ are periods of MFS and sine
functions, respectively.

Transmission of the periodically phase modulated field through the GDD circuit
changes phases of its frequency components as follows
\begin{equation}
E^{T}_{M,S}(t)=E(t)\sum_{n=-\infty}^{\infty}A^{M,S}_{n}(\Delta)e^{-i2\pi
nt/T_{M,S} +i\alpha_{M,S} n^{2}},\label{Eq5}%
\end{equation}
where $\alpha_{M,S}=2\pi^{2}\beta_{2}L/T_{M,S}^{2}$, $\beta_{2}$ is the group
velocity dispersion, and $L$ is the circuit length.

According to the concept of chirp radar, to create short pulses from the
phase-modulated CW field, there must be a certain relation between the
parameter $\beta_{2}L$, which defines phase shifts of the spectral components,
and the frequency chirp rate. This statement applies to linear chirp. For
example, parabolic phase change, $\varphi_{P}=-Kt^{2}/2$ gives a linear chirp
with the constant rate $K$. Theoretical analysis \cite{Shakhmuratov24} shows
that the condition for generation of pulses with maximum amplitude or the
bunching condition is $-K\beta_{2}L=1$. For the sine phase modulation
(\ref{Eq2}), this condition is slightly different since the sine modulation
produces nonlinear chirp with the chirping rate $C(t)=\partial^{2}\varphi
_{S}(t)/\partial t^{2}$. Its extreme value with proper sign is
$C_{\mathrm{ext}}=-\Delta\pi/T$ when $t=T/2+2nT$ and the chirp is close to
linear, where $n$ is an integer. If we choose this value in the bunching
condition $C_{\mathrm{ext}}\beta_{2}L=1$, replacing $-K$ with $C_{\mathrm{ext}%
}$, as suggested in \cite{Kobayashi1988,Kim1996,Otsuji1996}, then optimal
value of the parameter $\alpha$ is $1/2\Delta$. In fact, the optimal value is
$\sim1.5$ times larger. In \cite{Komukai2005}, it was numerically derived the
following approximate condition
\begin{equation}
\alpha_{S}=\frac{2\pi^{2}}{30.45\Delta-12.56},\label{Eq6}%
\end{equation}
to obtain the maximum peak power of pulses for $1<\Delta<10$.
\begin{figure}[ptb]
\resizebox{0.4\textwidth}{!}{\includegraphics{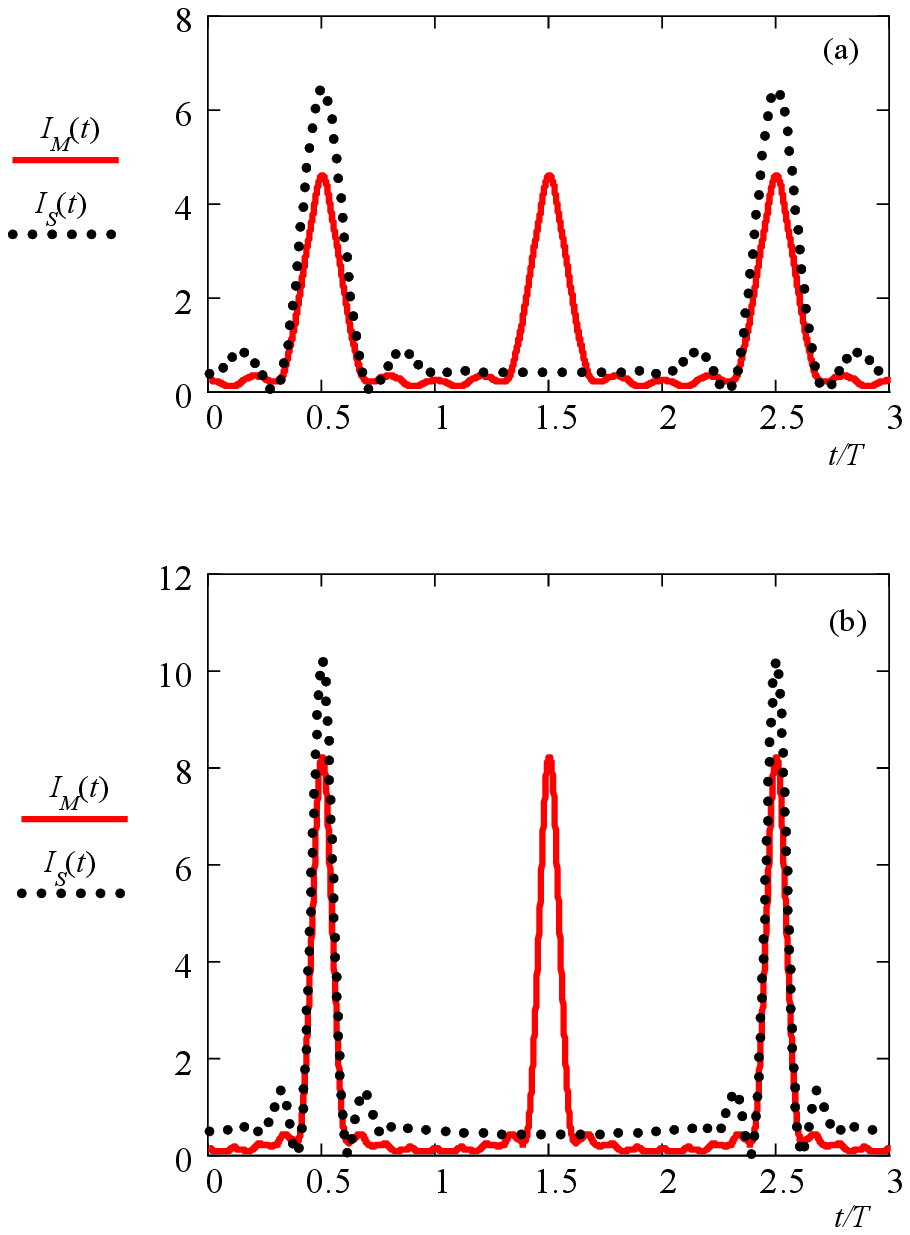}}\caption{Time
evolution of the intensity of the phase-modulated CW field, $I_{S,M}%
(t)=|E_{S,M}(t)|^{2}$, after propagating the GDD circuit with the optimal
value of the parameter $\alpha_{M,S}$. Sine modulated field is shown by black
dotted line for (a) $\Delta=\pi$ and $\alpha_{S}=0.25$, and (b) $\Delta=2\pi$
and $\alpha_{S}=0.11$. MFS modulated field is shown by red solid line for (a)
$\Delta=\pi$ and $\alpha_{M}=0.62$, and (b) $\Delta=2\pi$ and $\alpha
_{S}=0.34$. The field intensity is normalized to $I_{0}=|E_{0}|^{2}$.}%
\label{fig:2}%
\end{figure}

For the sinusoidal phase modulation (\ref{Eq2}), the amplitudes of the
spectral components are $A^{S}_{n}(\Delta)=J_{n}(\Delta)$, where $J_{n}%
(\Delta)$ is the Bessel function of the first kind and order $n$. Taking these
components and the optimal value of the parameter $\alpha_{S}$ in (\ref{Eq5}),
we can find the evolution of the filed over time that forms large pulses.
Examples of the pulses with maximum amplitudes obtained for $\Delta=\pi$ and
$\Delta=2\pi$ are shown in Fig. 2 by dotted black lines. It is evident that a
non-zero pedestal between pulses is formed. Noticeable sidelobes appear on the
wings of the pulses.

The spectrum of a CW field with phase modulation according to the MFS differs
significantly from the spectrum of the sinusoidal phase modulated field. The
amplitudes of the spectral components of the MFS modulated field will be
derived in the next section. Examples of pulses created by this field at
optimal values of the parameter $\alpha_{M}$ are shown in Fig. 2 by red solid
line. The pedestal between pulses is noticeably reduced, but a small ripple
appears on the wings.

One might expect that, due to coincidence of the time dependencies of
sinusoidal and MFS modulations in some time intervals, for example, between
$0$ and $T$ (see Fig. 1), there should be some simple relation between the
optimal values of the parameters $\alpha_{S}$ and $\alpha_{M}$ for the same
value of the modulation index. However, this is not the case, since the
spectra of sinusoidal and MFS modulated fields differ qualitatively.

\section{Spectrum of MFS-modulated field}

Since the functions $\varphi_{S}$ and $\varphi_{M}$ coincide on the time
interval $(0,T)$, it is possible to calculate the spectral components
$A^{M}_{n}(\Delta)$ in equation (\ref{Eq4}), taking into account that on this
interval the relation
\begin{equation}
e^{-i\varphi_{M}(t)}=e^{-i\varphi_{S}(t)} \equiv\sum_{n=-\infty}^{\infty}
J_{n}(\Delta)e^{-i2\pi n t/2T}\label{Eq7}%
\end{equation}
is valid, where the definitions of $T_{M}=T$ and $T_{S}=2T$ are used. The
result of analytical calculation is
\begin{equation}
A^{M}_{n}(\Delta)=J_{2n}(\Delta)+i\frac{4}{\pi}\sum_{k=0}^{\infty}
\frac{(2k+1)J_{2k+1}(\Delta)}{4n^{2}-(2k+1)^{2}}.\label{Eq8}%
\end{equation}
\begin{figure}[ptb]
\resizebox{0.45\textwidth}{!}{\includegraphics{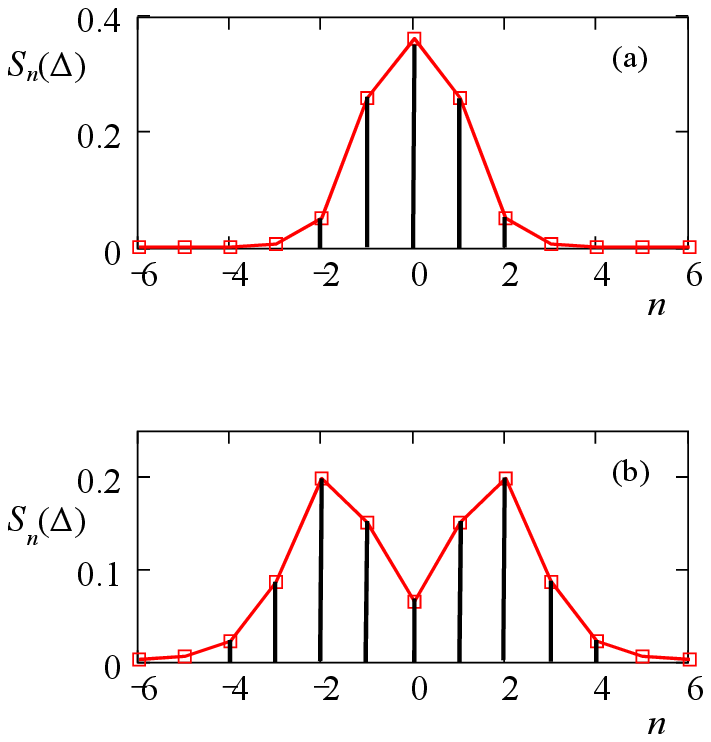}}.\caption{Non-dimensional
components of the field spectrum $S_{n}(\Delta)=|A^{M}_{n}(\Delta)|^{2}$ are
shown by black vertical bars for $\Delta=\pi$ (a) and $\Delta=2\pi$ (b). Red
lines connect the tips of the bars marked by red boxes for visualization.}%
\label{fig:3}%
\end{figure}Two examples of the field spectrum are shown in Fig. 3. For
$\Delta=\pi$ the spectrum can be approximated by a Gaussian. At $\Delta=2\pi$
a dip appears in the center of the spectrum. This dip starts to develop at
values of $\Delta> 3.5$.

The phases $\phi_{n}(\Delta)$ of the spectral components $A^{M}_{n}%
(\Delta)=\exp[i\phi_{n}(\Delta)]|A^{M}_{n}(\Delta)|$ have a peculiar
dependence on the component number $n$. If we compensate these phases, for
example, using liquid-crystal modulators
\cite{Weiner2006,Finot2019,Sheveleva2020,Sheveleva2021} that bring all
spectral components in phase, i.e.,
\begin{equation}
E^{P}_{M}(t)= E(t)\sum_{n=-\infty}^{\infty}|A^{M}_{n}(\Delta)|e^{-i2\pi
nt/T},\label{Eq9}%
\end{equation}
then Fourier-limited pulses are formed. Examples of these pulses,
$I_{P}(t)=|E^{P}_{M}|^{2}$, for $\Delta=\pi$ and $\Delta=\pi$ are shown in
Fig. 4(a). There are no sidelobes for $\Delta=\pi$ and the field decreases to
the level $-21.5$ dB between pulses. The shape of the pulses is close to the
secant hyperbolic pulse. Approximation of these pulses by $I_{sech}%
(t)=6.34I_{0} \text{sech} [(t-T)/(0.053T)]$ is shown in Fig. 4(b). At
$\Delta=2\pi$ the intensity of the pulses increases. Duration of the pulses
decreases by 2 times. Small sidelobes appear on the wings of the pulses. The
field intensity decreases to the level $-37$ dB between pulses. It can be
shown that the sidelobes appear due to the dip in the center of the field
spectrum. \begin{figure}[ptb]
\resizebox{0.4\textwidth}{!}{\includegraphics{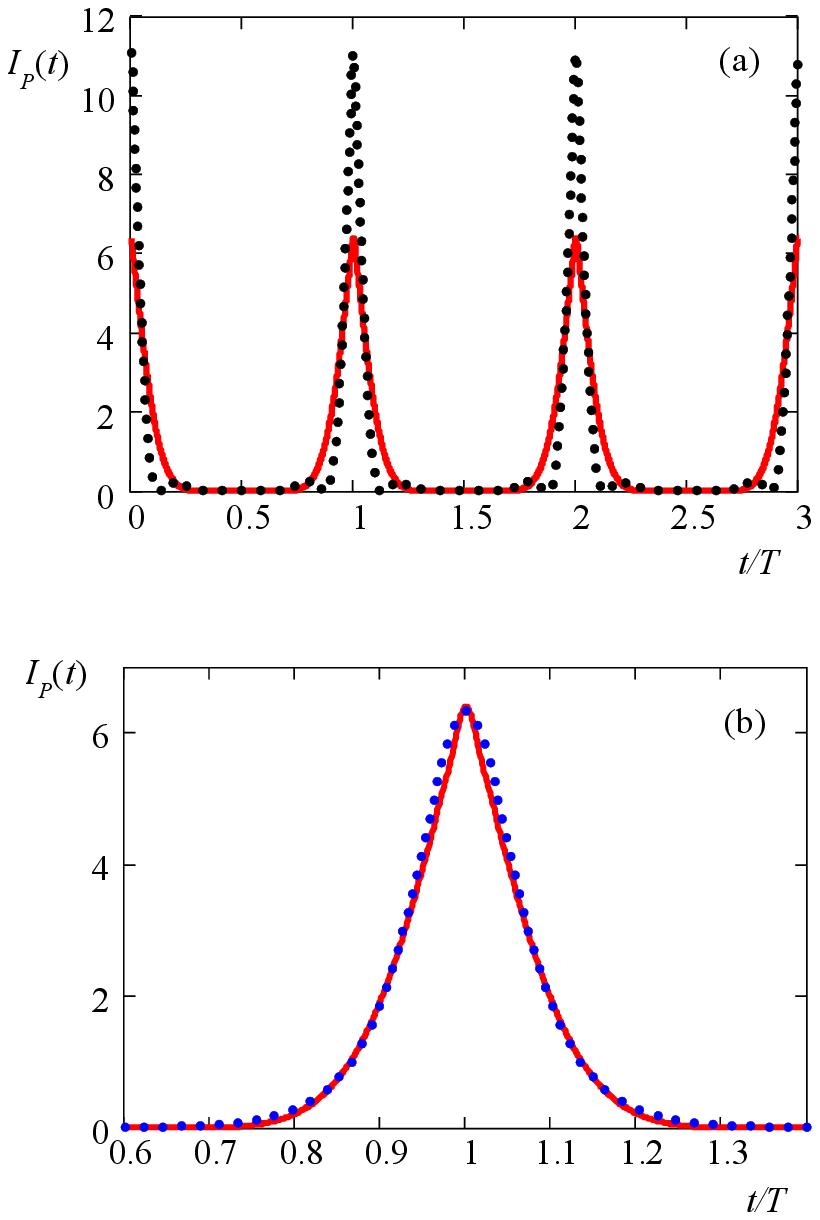}}.\caption{(a) Time
evolution of the field intensity of the MFS modulated field after phasing of
its spectral components for $\Delta=\pi$ (red solid line) and $\Delta=2\pi$
(black dotted line). (b) Approximation of the pulse (red line) by secant
hyperbolic pulse shape (dotted blue line) for $\Delta=\pi$. Intensity is
normalized to $I_{0}$.}%
\label{fig:4}%
\end{figure}

In the examples under consideration, the MFS phase-modulated field after phase
correction of the spectral components experiences their bunching into pulses
at $t=nT$, i.e., every time when all exponents in (\ref{Eq9}) are $\exp(i2\pi
nt/T)=1$, where $n$ is an integer. Therefore, the pulses are Fourier-limited.
The sinusoidal phase modulated field is formed into pulses at a suitable value
of $\alpha_{S}$ when $t=(T/2)+2Tn$, i.e., whenever all time exponents in
(\ref{Eq5}) are $\exp(i2\pi nt/2T)=(i)^{n}$. These are precisely those moments
in time when the chirping rate has the appropriate sign, and its absolute
value is maximum. Here we again take $T_{S}=2T$. Intensity of the pulses
increases with increasing modulation index $\Delta$ for both types of phase
modulation. These dependencies are shown in Fig. 5, demonstrating their
similarity. \begin{figure}[t]
\resizebox{0.4\textwidth}{!}{\includegraphics{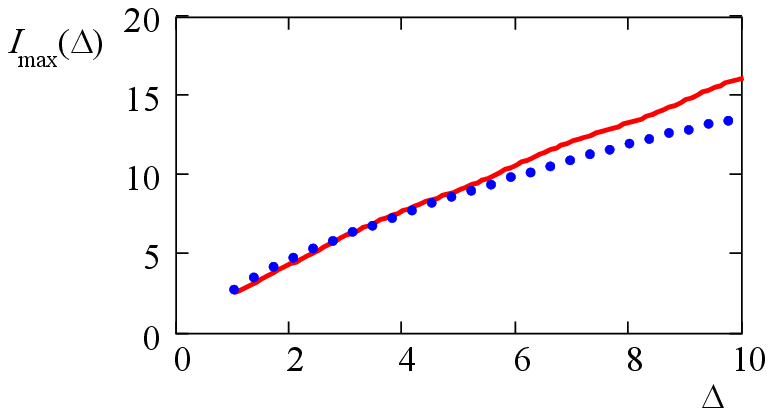}}\caption{Dependencies
of the maximum pulse intensity $I_{\max} (\Delta)$ on the value of modulation
index $\Delta$ for the MFS phase-modulated field (red solid line) and for the
sine phase-modulated field (blue dotted line). Parameter $\alpha_{S}$ for the
latter is taken from (\ref{Eq6}). Pulse intensity is normalized to $I_{0}$.}%
\label{fig:5}%
\end{figure}

To make all spectral components of the the MFS phase-modulated field have the
same phase, we introduce the phase shift $\psi_{n}(\Delta)=\pi/2 - \phi
_{n}(\Delta)$ of these components, which transforms them to $A^{M}_{n}%
(\Delta)\exp[i\psi_{n}(\Delta)]=\exp[i\pi/2]|A^{M}_{n}(\Delta)|$. Here the
common phase shift $\pi/2$ of the spectral components does not affect the time
evolution of the field intensity. The dependence of the necessary phase shift
$\psi_{n}(\Delta)$ on the number $n$ is shown in Fig. 6. It is obvious that
for large $\pm n$ this phase tends to zero. The dependence of the phase
$\psi_{n}(\Delta)$ on $n$ is close to Lorentzian. Similar but different phase
shifts have been proposed to create "besselon" from a sinusoidal
phase-modulated field by triangular spectral phase tailoring, which generates
high-repetition-rate picosecond pulse trains \cite{Finot2019,Sheveleva2021}.

If we limit the domain of definition of the phase $\psi_{n}(\Delta)$ to the
interval $(-\pi,\pi)$, then for $\Delta=\pi$ we get $\psi_{0}(\pi)=-0.831\pi$,
which differs from that shown in Fig. 6 by $-2\pi$. This phase offset does not
affect on the value of $\exp[i\psi_{0}(\pi)]$. For $\Delta=2\pi$ the phases of
three spectral components are redefined as $\psi_{0}(2\pi)=0.33\pi$ and
$\psi_{\pm1}(2\pi)=-0.735\pi$. \begin{figure}[t]
\resizebox{0.4\textwidth}{!}{\includegraphics{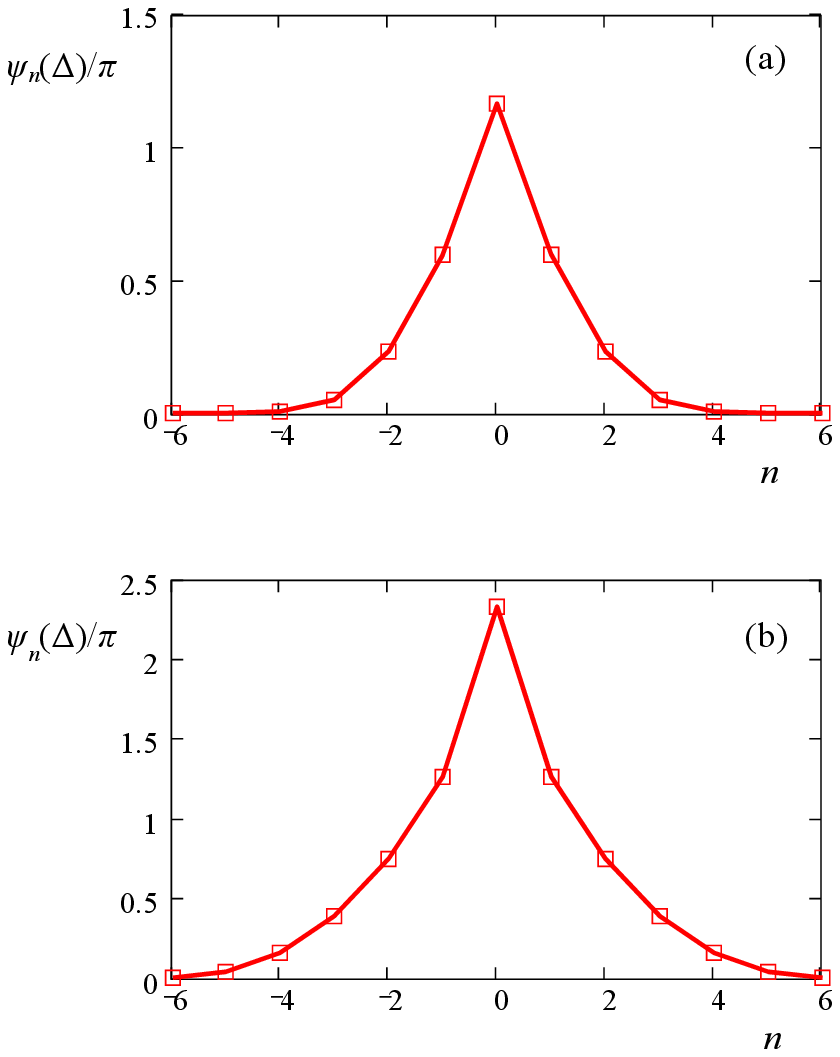}}.\caption{Dependence
of the necessary phase shifts of the spectral components of the MFS
phase-modulated field on the component number for $\Delta=\pi$ (a) and
$\Delta=2\pi$ (b).}%
\label{fig:6}%
\end{figure}

Because for large $n$ the tuning phase $\psi_{n}(\Delta)$ tends to zero, one
can limit the phase tailoring of the spectral components to a few numbers
around central part of the spectrum. Examples where for $\Delta=\pi$ only the
phases of spectral components $0$, $\pm1$, and $\pm2$ are manipulated are
shown in Fig. 7(a). The phase tailoring accuracy of $0.1$ rad is sufficient.
For $\Delta=2\pi$ we need to introduce the phase change also in the $\pm3$
spectral components, see Fig. 7(b). \begin{figure}[t]
\resizebox{0.45\textwidth}{!}{\includegraphics{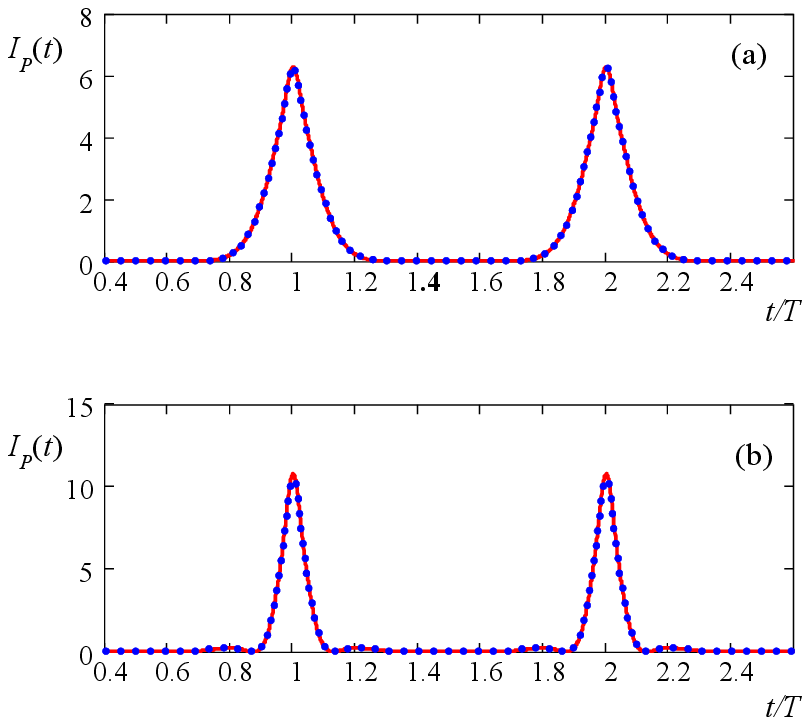}}\caption{Time
evolution of the field intensity after phase tailoring of its spectral
components (normalized to $I_{0}$). Modulation index is $\Delta=\pi$ in (a)
and only 5 spectral components are phase manipulated (dotted blue line). The
phase correction values are $\psi_{0}(\pi)=2.6$ rad, $\psi_{\pm1}(\pi)=1.9$
rad, and $\psi_{\pm2}(\pi)=0.7$ rad. Modulation index is $\Delta=2\pi$ in (b)
and only 7 spectral components are phase manipulated (dotted blue line). The
phase correction values are $\psi_{0}(2\pi)=1$ rad, $\psi_{\pm1}(2\pi)=-2.3$
rad, $\psi_{\pm2}(2\pi)=2.3$ rad, and $\psi_{\pm3}(2\pi)=1.2$ rad. The cases
when all spectral components are phase tailored are shown by red solid line.}%
\label{fig:7}%
\end{figure}\begin{figure}[tt]
\resizebox{0.4\textwidth}{!}{\includegraphics{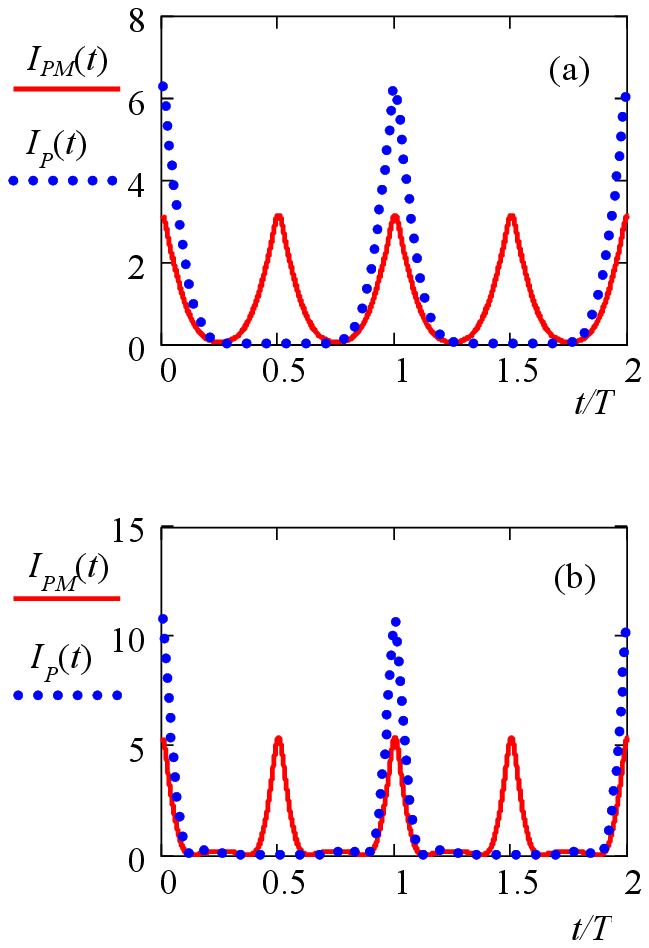}}.\caption{Comparison
of pulses produced by phase tailoring of spectral components of the MFS
phase-modulated field (blue dots) with those obtained after additional $\pi/2$
phase shifts of odd spectral components (red solid line). Modulation index is
$\Delta=\pi$ (a) and $\Delta=2\pi$ (b).}%
\label{fig:8}%
\end{figure}

\section{Multiplying repetition rate of the pulses}

The pulse repetition rate can be increased by a factor of $N$, where $N$ is an
integer, by transmitting the pulses through a suitable GDD circuit with the
required fractional Talbot effect or through an unbalanced Mach-Zehnder
interferometer with $N$ arms with specially selected phase shifts and delay
times in each arm \cite{Berger2003,Berger2004}. These methods make it possible
to multiply the repetition rate of a periodic sequence of pulses without
distorting the individual characteristics of the pulses. For $N=2$ the
suitable GDD circuit transforms the field $E^{P}_{M}(t)$ in Eq. (\ref{Eq9})
to
\begin{equation}
E_{PM}(t)= E(t)\sum_{n=-\infty}^{\infty}|A^{M}_{n}(\Delta)|e^{-i2\pi nt/T+
i\pi n^{2}/2}.\label{Eq10}%
\end{equation}
This GDD circuit satisfies $1/Q$-fractional Talbot effect with $Q=4$, where
$1/Q=\pi\beta_{2}L/T_{M}^{2}$. It can be shown
\cite{Shakhmuratov24,Shakhmuratov22,Shakhmuratov22LP} that (\ref{Eq10}) is
reduced to
\begin{equation}
E_{PM}(t)=\frac{e^{i\frac{\pi}{4}}E^{P}_{M}(t)+e^{-i\frac{\pi}{4}}E^{P}%
_{M}(t+\frac{T}{2})}{\sqrt{2}}.\label{Eq11}%
\end{equation}
This is the sum of two time- and phase-shifted replicas of the field
$E^{P}_{M}(t)$, which can be obtained using unbalanced Mach-Zehnder
interferometer with two arms.

Both equations, (\ref{Eq10}) and (\ref{Eq11}), can be reduced to $E_{PM}(t)=
E_{1}(t)+E_{2}(t)$, where
\begin{equation}
E_{1}(t)= E(t) \sum_{n=-\infty}^{\infty} |A^{M}_{2n}(\Delta)| e^{-i2\pi
2nt/T}\label{Eq12}%
\end{equation}
is a sum of even spectral components and
\begin{equation}
E_{2}(t)= E(t)\sum_{n=-\infty}^{\infty}|A^{M}_{2n+1}(\Delta)| e^{-i2\pi
(2n+1)t/T+i\pi/2}\label{Eq13}%
\end{equation}
is a sum of odd spectral components shifted in phase by $\pi/2$. Therefore, by
introducing $\pi/2$-phase shifts of odd spectral components using
liquid-crystal modulators, it is also possible to increase the pulse
repetition rate by 2 times.

Examples of repetition rate multiplication are shown in Fig. 8. The maximum
pulse intensity is halved as the field is redistributed between the two
replicas. The extinction ratio, defined as the ratio of the maximum to the
minimum value of the pulse intensity, is 18.3 dB for $\Delta=\pi$ and 30.7 dB
for $\Delta=2\pi$. However, there is a small pedestal between the pulses for
$\Delta=2\pi$, and the ratio of the pulse maximum to the maximum of this
pedestal is 15.4 dB.

\section{Additive synthesis of MFS-modulation with several harmonics}

A high-frequency oscillating-voltage rectifier is not yet available.
Therefore, to create the proposed MFS phase modulation of the field by the
electro-optical modulator, it is possible to use a multi-frequency oscillating
voltage. Some examples of multi-tone microwave signals modulated on the
optical field are considered in \cite{Zhang2024}. \begin{figure}[t]
\resizebox{0.4\textwidth}{!}{\includegraphics{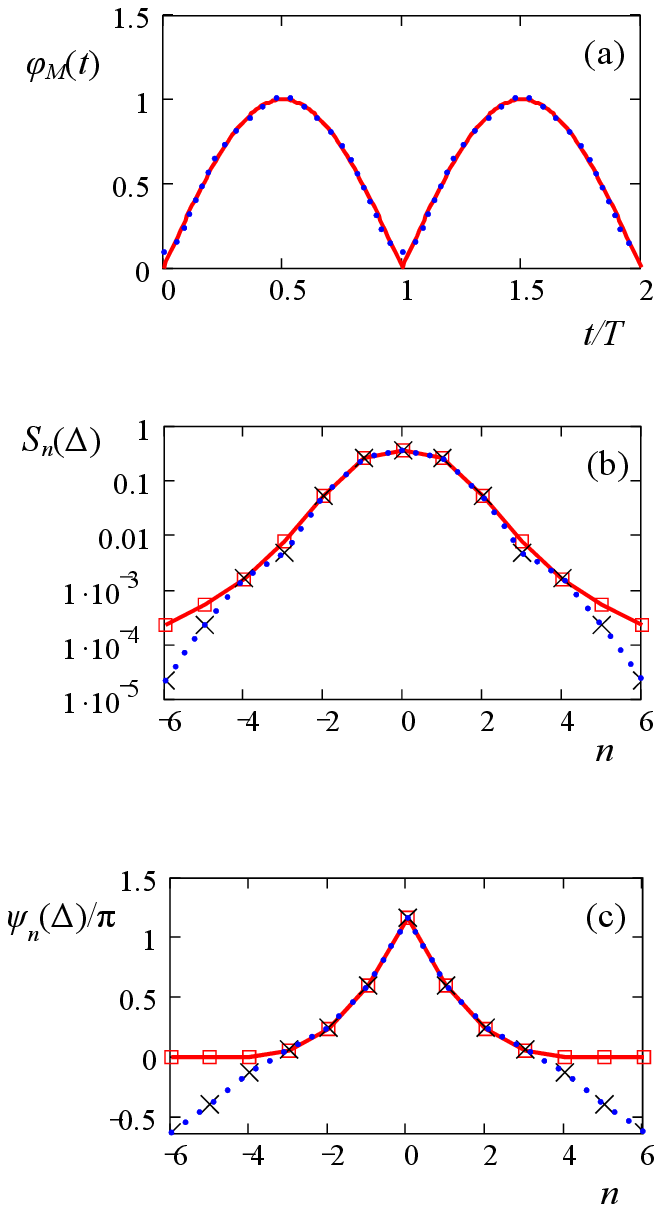}}.\caption{(a) SM
phase modulation (red solid line) and three-tone phase modulation (dotted blue
line). Both are normalized to $\Delta$. (b) Non-dimensional components of the
field spectrum of the SM phase modulated field (red boxes connected by red
lines) and of the three-tone phase modulation (blue crosses connected by blue
dotted lines) for $\Delta= \pi$. (c) Phase values required for phasing the
spectral components of the SM phase-modulated field (red boxes connected by
red lines) and the spectral components of the three-tone phase modulated field
(blue crosses connected by blue dotted lines) for $\Delta= \pi$.}%
\label{fig:9}%
\end{figure}\begin{figure}[tt]
\resizebox{0.4\textwidth}{!}{\includegraphics{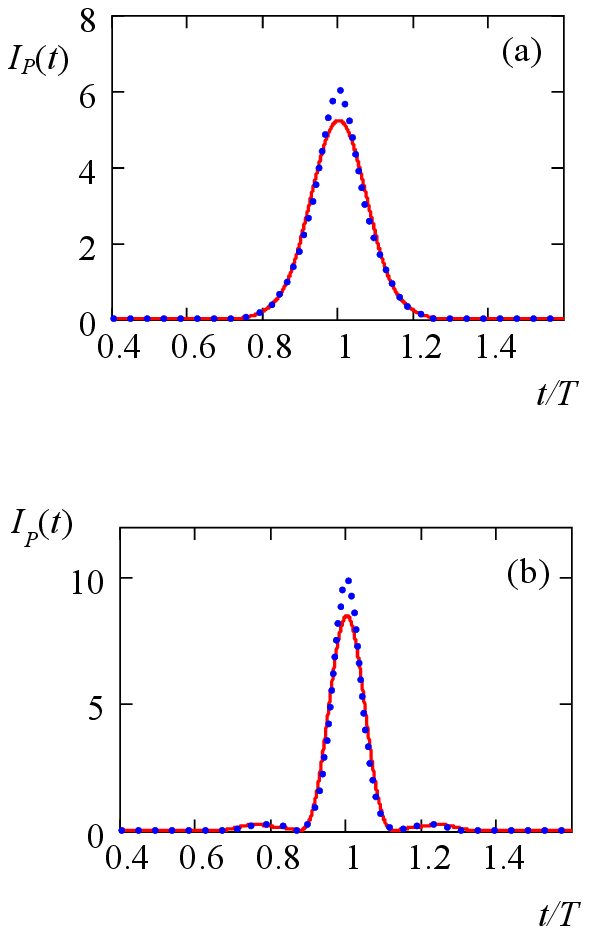}}.\caption{Pulses
produced by phasing a limited number of spectral components of the three-tone
phase-modulated field (red solid line) for the modulation index values
$\Delta=\pi$ (a) and $\Delta=2\pi$ (b). For comparison, the blue dotted line
shows pulses obtained by phasing the same number of spectral components of the
SM phase-modulated field.}%
\label{fig:10}%
\end{figure}

Below we take a Fourier transform of MFS modulation, which is
\begin{equation}
\varphi_{M}(t)=\frac{2\Delta}{\pi}- \frac{4\Delta}{\pi}\sum_{n=1}^{N}
\frac{\cos(2\pi n t/T )}{4n^{2}-1},\label{Eq14}%
\end{equation}
where $N$ is the number of harmonics, which we limit. A comparison of the MFS
phase modulation with (\ref{Eq14}), where we limit the number of harmonics to
$N=3$, is shown in Fig. 9(a). This limitation leads to smoothing of the sharp
change in the phase evolution around time $t=T$, which is clearly visible in
Fig. 9(a). The numerically calculated intensity of the spectral components
$S_{n}(\Delta)=|A^{M}_{n}(\Delta)|^{2}$ of the field with a given phase
modulation is compared with the spectrum of MFS phase-modulated field in Fig.
9(b) for $\Delta=\pi$. The central spectral components (up to $\pm4$) coincide
quite well with those of the MFS phase modulated field. Then the higher
spectral components of the three-tone phase-modulated field fall off much
faster. Comparison of phases compensating the phases of the spectral
components of the field and making them in-phase for the MFS phase-modulated
field and three-tone phase-modulated field is shown in Fig. 9 (c). Similar
dependencies can be found for $\Delta=2\pi$.

For three-tone phase modulation, examples of pulses obtained by phasing the
five central spectral components for $\Delta=\pi$ and the seven central
spectral components for $\Delta=2\pi$ are shown in Fig. 10 by red solid lines.
They are compared with pulses obtained by phasing the same number of spectral
components of the MFS phase-modulated field (shown by dotted blue line). With
three-tone phase modulation, the pulse shape is closer to Gaussian. The peak
intensity of the pulses is reduced due to a faster drop of the spectral wings
of the three-tone phase-modulated field compared to the spectrum of the field
created by MFS phase modulation, see Fig. 9(b).

The advantage of three-tone phase modulation is a significant reduction of the
required modulation index. If we define this index in terms of half-wave
voltage $V_{\pi}$ that produces the $\pi$-phase shift of the field, then, for
example, for $\Delta=2\pi$ we need $0.85V_{\pi}$ to generate the first
harmonic, $0.17V_{\pi}$ for the second harmonic, and $0.07V_{\pi}$ for the
third harmonic. The constant phase shift $2\Delta/\pi$ in (\ref{Eq14}) can be
omitted since it is the same for all spectral components of the
phase-modulated field.

\section{Conclusion}

A multi-tone phase and amplitude modulation, composed of a number of
sinusoidal signals with specific frequencies and amplitudes, allows generating
pulses of arbitrary shape using the fractional Talbot effect
\cite{Zhang2024,Chi2021}. Multi-tone phase-only modulation, simulating
periodic sawtooth, rectangular or parabolic phase evolution, also allows pulse
generation using the fractional Talbot effect or by removing the central
component of the frequency comb produced by phase modulation
\cite{Shakhmuratov24,Shakhmuratov22,Shakhmuratov23}. In this paper it is shown
that another phase tailoring of the spectrum, obtained using the MFS phase
modulation, leads to the generation of pulses. This spectrum considerably
differs from the spectrum of the sinusoidal phase modulated CW field. The
phases of the spectral components are symmetrically distributed around the
central component. This kind of spectrum can be obtained by three-tone phase
modulation. The produced pulses have no noticeable pedestal. By choosing a
moderate value of the modulation index, it is possible to generate pulses
without sidelobes. Using the same procedure of phase tailoring, which includes
additional $\pi/2$-phase shifts of odd spectral components, the pulse
repetition rate can be doubled. Three-tone phase modulation allows to
significantly reduce the modulation indices of each tone required to obtain
the desired pulses. The proposed method allows to generate picosecond pulses
with a repetition rate of 20 GHz and a short time interval between pulses if
the fundamental tone has a frequency of 10 GHz.

\end{document}